\documentclass[final,5p,times,twocolumn,nopreprintline]{elsarticle}
\pdfoutput=1

\usepackage{amsmath,slashed,booktabs}
\usepackage{graphicx,graphics}
\usepackage{dcolumn}
\usepackage[hyperfootnotes=false]{hyperref}
\usepackage{xspace}
\usepackage{color}
\usepackage{balance}
\usepackage{subcaption}
\usepackage{multirow}
\usepackage{multicol}
\usepackage{cancel}

\usepackage{fancyhdr}
\fancypagestyle{firstpage}{%
	
	\lhead{}
	\rhead{\small
	PSI-PR-23-32,
	ZU-TH 48/23}
}
\newcommand{\E}{\mathcal{E}}
\renewcommand{\L}{\mathcal{L}}
\newcommand{\M}{\mathcal{M}}
\newcommand{\N}{\mathcal{N}}
\renewcommand{\O}{\mathcal{O}}
\newcommand{\tr}{\mathrm{Tr}}
\newcommand{\p}{\partial}
\newcommand{\cptgo}{\mbox{\cancel{CP}-3GO}}
\newcommand{\nn}{\nonumber\\}
\newcommand{\<}{\langle}
\renewcommand{\>}{\rangle}

\allowdisplaybreaks[1]

\begin{document}

\renewcommand{\theequation}{\arabic{equation}}

\begin{frontmatter}
 
\title{One-loop matching of the $CP$-odd three-gluon operator to the gradient flow}

\author[UZH,PSI]{Òscar L.~Crosas}

\author[WM]{Christopher J.~Monahan}

\author[MSU]{Matthew~D.~Rizik}

\author[Aachen]{Andrea Shindler}

\author[UZH,PSI]{Peter Stoffer}

\address[UZH]{Physik-Institut, Universit\"at Z\"urich, Winterthurerstrasse 190, 8057 Z\"urich, Switzerland}
\address[PSI]{Paul Scherrer Institut, 5232 Villigen PSI, Switzerland}
\address[WM]{Department of Physics, The College of William \& Mary, Williamsburg, VA 23187, USA}
\address[MSU]{Facility for Rare Isotope Beams \& Physics Department, Michigan State University, East Lansing, Michigan 48824, USA}
\address[Aachen]{Institute for Theoretical Particle Physics and Cosmology, RWTH Aachen University, 52056 Aachen, Germany}

\begin{abstract}
The calculation of the neutron electric dipole moment within effective field theories for physics beyond the Standard Model requires non-perturbative hadronic matrix elements of effective operators composed of quark and gluon fields. In order to use input from lattice computations, these matrix elements must be translated from a scheme suitable for lattice QCD to the minimal-subtraction scheme used in the effective-field-theory framework. The accuracy goal in the context of the neutron electric dipole moment necessitates at least a one-loop matching calculation. Here, we provide the one-loop matching coefficients for the $CP$-odd three-gluon operator between two different minimally subtracted 't~Hooft--Veltman schemes and the gradient flow. This completes our program to obtain the one-loop gradient-flow matching coefficients for all $CP$-violating and flavor-conserving operators in the low-energy effective field theory up to dimension six.
\end{abstract}

\end{frontmatter}

\thispagestyle{firstpage}


\section{Introduction}
\label{sec:Introduction}

The baryon asymmetry of the universe calls for $CP$ violation beyond the Standard Model (SM) of particle physics, which however is tightly constrained by the experimental bounds on the neutron electric dipole moment (EDM)~\cite{Abel:2020pzs}
\begin{align}
	\label{eq:nEDMExperimentalBound}
	|d_n| < 1.8 \times 10^{-26} \, e\, \mathrm{cm} \text{ (90\% C.L.)} \, ,
\end{align}
as well as on leptonic~\cite{Muong-2:2008ebm,ACME:2018yjb,Adelmann:2021udj,Roussy:2022cmp} and other hadronic EDMs, see Refs.~\cite{Chupp:2017rkp,Shindler:2021bcx,Alarcon:2022ero} for recent reviews. Effective field theories (EFTs) are the ideal framework to translate the experimental bounds at low energies into constraints on heavy new physics beyond the SM. Between the threshold of new physics and the electroweak scale, the effects of new physics can be calculated in terms of the SMEFT~\cite{Buchmuller:1985jz,Grzadkowski:2010es,Jenkins:2013zja,Jenkins:2013wua,Alonso:2013hga}, which below the electroweak scale is matched to the low-energy effective field theory (LEFT)~\cite{Jenkins:2017jig,Jenkins:2017dyc,Dekens:2019ept}. Within the LEFT, the neutron EDM depends on non-perturbative matrix elements of composite operators. These matrix elements should be provided with a precision of $10-25\%$~\cite{Alarcon:2022ero} in order not to wash out the constraining power of the low-energy experimental bound when translating that bound into constraints on the Wilson coefficients at the scale of heavy new physics. Ideally, lattice QCD should be used to determine these matrix elements, which, however, requires a matching calculation between a non-perturbative renormalization scheme and minimal subtraction (MS) used in the EFTs. These matching calculations are partially available for traditional momentum-subtraction schemes~\cite{Bhattacharya:2015rsa,Cirigliano:2020msr}, but on the lattice side these schemes come with the complication of large operator bases and power divergences that interfere with the continuum limit. The gradient flow~\cite{Luscher:2010iy,Luscher:2011bx,Luscher:2013cpa} (see Ref.~\cite{Shindler:2022tlx} for a recent review) is a more modern scheme that promises to give a handle on these difficulties~\cite{Rizik:2020naq,Kim:2021qae}.

Lattice-QCD computations of the matrix elements of flowed operators can be converted into the required matrix elements of MS operators by using a perturbative matching at a renormalization scale that is accessible on the lattice and where perturbation theory still works sufficiently well.

The perturbative matching equations between the gradient flow and the MS scheme have recently been worked out at one loop for all operators contributing to the neutron EDM up to dimension six~\cite{Rizik:2020naq,Mereghetti:2021nkt,Buhler:2023gsg}, with the exception of the $CP$-odd three-gluon operator (\cptgo).\footnote{This operator is already part of the SMEFT operator basis of Ref.~\cite{Buchmuller:1985jz}, but it is often called ``Weinberg operator''~\cite{Weinberg:1989dx}, or gluon chromo-EDM~\cite{Braaten:1990zt,Cirigliano:2020msr}.} The matching of this operator is the most involved calculation, due to the number of operators that can mix with the \cptgo{}~\cite{Cirigliano:2020msr} and due to the complexity of the gluonic diagrams. In this letter, we present the matching for the \cptgo{} at one loop, which completes the one-loop gradient-flow matching up to dimension six for the flavor-conserving and $CP$-violating sector of the LEFT.


\section{Gradient flow}
\label{sec:Flow}

The gradient flow~\cite{Luscher:2010iy,Luscher:2011bx,Luscher:2013cpa} extends Euclidean QCD
\begin{equation}
	\label{eq:EuclideanQCD}
	\L_\mathrm{QCD} = \frac{1}{4g_0^2} G_{\mu\nu}^a G_{\mu\nu}^a + \bar q(\slashed D + \M) q \, , \quad D_\mu = \p_\mu + t^a G_\mu^a
\end{equation}
to $D+1$ dimensions by introducing an artificial dimension called flow time $t$, with mass dimension $[t] = -2$. In the following, we use the same conventions as Refs.~\cite{Mereghetti:2021nkt,Buhler:2023gsg}. The flowed fields satisfy the flow equations
\begin{align}
	\label{eq:FlowEquations}
	\p_t B_\mu &= D_\nu G_{\nu\mu} + \alpha_0 D_\mu \p_\nu B_\nu \, , \nn
	\p_t \chi &= D_\mu D_\mu \chi - \alpha_0 (\p_\mu B_\mu) \chi \, , \nn
	\p_t \bar\chi &= \bar\chi \overleftarrow D_\mu \overleftarrow D_\mu + \alpha_0 \bar\chi (\p_\mu B_\mu)
\end{align}
together with the initial conditions
\begin{align}
	B_\mu(x,t=0) &= G_\mu(x) \, , \nn
	\chi(x,t=0) &= q(x) \, , \quad \bar\chi(x,t=0) = \bar q(x) \, ,
\end{align}
i.e., the theory is defined as QCD at the boundary $t=0$.
The flow equations can be extended to include an external electromagnetic field in order to preserve $U(1)_\mathrm{em}$ invariance~\cite{Buhler:2023gsg}. In the present case, up to one loop there is no matching to physical operators containing photons, hence we disregard the inclusion of an external photon field.

The differential equations~\eqref{eq:FlowEquations} can be rewritten as integral equations and solved perturbatively in an expansion in the gauge coupling~\cite{Luscher:2011bx}. This expansion is conveniently expressed in terms of Feynman diagrams with QCD Feynman rules extended by flow lines and flow vertices.

The gradient flow acts as a UV regulator: the flowed theory is automatically finite, apart from the renormalization of the gauge coupling, quark masses, and quark fields. In particular, flowed composite operators are renormalized only multiplicatively by factors of the coupling, mass, and quark-field renormalizations.

When implemented non-perturbatively on the lattice, the gradient flow converts power divergences into $1/t^n$ singularities, disentangling them from the continuum limit $a\to0$, which can be taken for any fixed flow time $t$.


\section{Operator basis}
\label{sec:Operators}

We largely follow the 't~Hooft--Veltman (HV)~\cite{Breitenlohner:1977hr,tHooft:1972tcz} scheme definitions of Ref.~\cite{Cirigliano:2020msr}, adapting them in a minimal way to the Euclidean conventions of Refs.~\cite{Mereghetti:2021nkt,Buhler:2023gsg} and rescaling the gauge field as in Eq.~\eqref{eq:EuclideanQCD}. In the HV scheme, the $D$-dimensional metric tensor $\delta_{\mu\nu}$ is split into a four-dimensional part $\bar\delta_{\mu\nu}$ and a part $\hat\delta_{\mu\nu}$ projecting to $-2\varepsilon$ dimensions,
\begin{equation}
	\delta_{\mu\nu} = \bar\delta_{\mu\nu} + \hat\delta_{\mu\nu} \, ,
\end{equation}
with
\begin{equation}
	\delta_{\mu\nu} \delta_{\mu\nu} = D = 4 - 2\varepsilon \, , \quad \bar\delta_{\mu\nu} \bar\delta_{\mu\nu} = 4 \, , \quad \hat\delta_{\mu\nu} \hat\delta_{\mu\nu} = -2\varepsilon \, .
\end{equation}
Projections of Dirac matrices, vectors, and tensors to the different sub-spaces are defined by contractions with the respective metric tensors, e.g.,
\begin{equation}
	\bar\gamma_\mu = \bar\delta_{\mu\nu} \gamma_\nu \, , \quad \hat\gamma_\mu = \hat\delta_{\mu\nu} \gamma_\nu \, .
\end{equation}
The Levi-Civita symbol $\epsilon_{\mu\nu\lambda\sigma}$ is a purely four-dimensional object and we use the convention
\begin{equation}
	\gamma_5 = \gamma_1 \gamma_2 \gamma_3 \gamma_4 = \frac{1}{4!} \epsilon_{\mu \nu \lambda \sigma} \gamma_\mu \gamma_\nu \gamma_\lambda \gamma_\sigma \, , \quad \epsilon_{1 2 3 4}=+1 \, .
\end{equation}
We define the renormalized flowed \cptgo{} as
\begin{equation}
	\label{eq:FlowedCP3GO}
	\O^R_{\widetilde G}(x,t) \coloneqq \frac{1}{g^2} \tr[G_{\mu\nu} G_{\nu\lambda} \widetilde G_{\lambda\mu}]
 \end{equation}
with renormalized coupling $g$ and the dual field-strength tensor $\widetilde{G}_{\mu\nu}\coloneqq \frac{1}{2}\varepsilon_{\mu\nu\rho\sigma}G_{\rho\sigma}$ defined in terms of flowed gauge fields.\footnote{We take the renormalized gauge coupling $g$ in the MS scheme.
In lattice simulations, an alternative gauge coupling definition is employed. 
This scheme change for the coupling can be easily implemented in the one-loop \cptgo{} self-matching~\cite{Luscher:2010iy,Harlander:2016vzb}, while at the order we work in this paper, the distinction holds no significance for the matching on the remaining operators.}
An operator-product expansion at short flow times (SFTE) expresses it in terms of renormalized MS operators
\begin{align}
	\O^R_{\widetilde G}(x,t) &= \sum_i C_i(t,\mu) \O_i^\mathrm{MS}(x,\mu) + \sum_i C_{\N_i}(t,\mu)\N_i^\mathrm{MS}(x,\mu) \nn
		&\quad + \sum_i C_{\E_i}(t,\mu)\E_i^\mathrm{MS}(x,\mu) \,,
\end{align}
where $\O_i$ denote physical operators (also called class-I operators), $\N_i$ nuisance operators (or class-II operators), and $\E_i$ evanescent ones, which vanish in four space-time dimensions.
The complete operator basis has been derived in Ref.~\cite{Cirigliano:2020msr}. Again, we minimally adapt the operator definitions to our conventions and we only include operators relevant at one loop. The physical operators $\O_i$ are defined as
\begin{align}
	\O_{\theta} &= \frac{1}{g_0^2} \tr[G_{\mu\nu} \widetilde G_{\mu\nu}] \, , \nn
	\O_{\widetilde G} &= \frac{1}{g_0^2} \tr[G_{\mu\nu} G_{\nu\lambda} \widetilde G_{\lambda\mu}] \, , \nn
	\O_{CE} &=  ( \bar q \tilde\sigma_{\mu\nu} \M t^a q ) G_{\mu\nu}^a \, , \nn
	\O_{\p G} &= \frac{1}{g_0^2} \p_\nu \tr[ (D_\mu G_{\mu\lambda}) \widetilde G_{\nu\lambda}] \, , \nn
	\O_{\Box\theta} &= \frac{1}{g_0^2} \Box \, \tr[G_{\mu\nu} \widetilde G_{\mu\nu}] \, ,
\end{align}
in terms of bare unflowed fields, bare coupling $g_0$, and with $\tilde\sigma_{\mu\nu} = -\frac{1}{2} \epsilon_{\mu\nu\alpha\beta} \sigma_{\alpha\beta}$. The nuisance operators $\N_i$ can be split into two classes~\cite{Dixon:1974ss,Kluberg-Stern:1975ebk,Joglekar:1975nu,Deans:1978wn,Collins:1984xc}. The gauge-invariant class-IIa operators vanish by the classical equations of motion (EOM):
\begin{align}
	\N_1 &= \left( \bar q_E \tilde\sigma_{\mu\nu} t^a q + \bar q \tilde\sigma_{\mu\nu} t^a q_E \right) G_{\mu\nu}^a \, , \nn
	\N_3 &= \bar q_E \M \tilde\gamma_\mu D_\mu q + \bar q \overleftarrow D_\mu \tilde\gamma_\mu \M q_E \, , \nn
	\N_6 &= \p_\mu \left( \bar q_E \gamma_5 D_\mu q - \bar q \overleftarrow D_\mu \gamma_5 q_E \right) \, , \nn
	\N_7 &= \p_\mu \left( \bar q_E \tilde\sigma_{\mu\nu} D_\nu q - \bar q \overleftarrow D_\nu \tilde\sigma_{\mu\nu} q_E \right) \, , \nn
	\N_8 &= \frac{2}{g_0^2} \p_\lambda\left( \widetilde G_{\lambda\sigma}^a \left( D_\rho G_{\rho\sigma}^a - g_0^2 \bar q t^a \gamma_\sigma q \right) \right) \, , \nn
	\N_9 &= \p_\mu \left( \bar q_E \M \tilde\gamma_\mu q + \bar q \M \tilde\gamma_\mu q_E \right) \, ,
\end{align}
where the quark EOM fields are
\begin{equation}
	q_E = (\slashed D + \M) q \, , \quad \bar q_E = \bar q ( - \overleftarrow{\slashed D} + \M)
\end{equation}
and $\tilde\gamma_\mu = \bar\gamma_\mu \gamma_5 = \frac{1}{3!} \epsilon_{\mu\alpha\beta\gamma} \gamma_\alpha \gamma_\beta \gamma_\gamma$.
Class IIb consists of gauge-variant, but BRST-exact operators:
\begin{align}
	\N_{11} &= \frac{2}{g_0^2}  \widetilde G_{\lambda\sigma}^a\left( \p_\lambda \left( D_\rho G_{\rho\sigma}^a - g_0^2 \bar q t^a \gamma_\sigma q + g_0^2 f^{abc} (\p_\sigma \bar c^b) c^c \right) \right) \, , \nn
	\N_{12} &= \left( \bar q_E \gamma_5 q + \bar q \gamma_5 q_E \right) G_\mu^a G_\mu^a \, , \nn
	\N_{13} &= \left( \bar q_E \gamma_5 t^a q + \bar q \gamma_5 t^a q_E \right) G_\mu^b G_\mu^c d^{abc} \, , \nn
	\N_{14} &= \left( \bar q_E \gamma_5 t^a q - \bar q \gamma_5 t^a q_E \right) \p_\mu G_\mu^a \, , \nn
	\N_{15} &= \left( \bar q_E \gamma_5 t^a D_\mu q - \bar q \overleftarrow D_\mu \gamma_5 t^a q_E \right) G_\mu^a \, , \nn
	\N_{16} &= \left( \bar q_E \tilde\sigma_{\mu\nu} t^a q + \bar q \tilde\sigma_{\mu\nu} t^a q_E \right) \p_\mu G_\nu^a \, , \nn
	\N_{17} &= \left( \bar q_E \M \tilde\gamma_\mu t^a q - \bar q \M \tilde\gamma_\mu t^a q_E \right) G_\mu^a \, , \nn
	\N_{18} &= \frac{\epsilon_{\mu\nu\lambda\sigma}}{g_0^2} \p_\lambda \left( \left( D_\rho G_{\rho\sigma}^a - g_0^2 \bar q t^a \gamma_\sigma q + g_0^2 f^{abc} (\p_\sigma \bar c^b) c^c \right) \p_\mu G_\nu^a \right) \, , \nn
	\N_{19} &= \p_\mu \left( \left( \bar q_E \gamma_5 t^a q - \bar q \gamma_5 t^a q_E \right) G_\mu^a  \right) \, , \nn
	\N_{20} &= \p_\mu \left( \left( \bar q_E \tilde\sigma_{\mu\nu} t^a q + \bar q \tilde\sigma_{\mu\nu} t^a q_E \right) G_\nu^a  \right) \, ,
\end{align}
which could be dropped from the operator basis when using the background-field method~\cite{Abbott:1980hw,Abbott:1983zw,Suzuki:2015bqa}.

Finally, the definition of the set of evanescent operators $\E_i$ affects the matching coefficients of the physical operators and is part of the scheme.\footnote{In a scheme that avoids a tree-level matching contribution to evanescent operators, their renormalization~\cite{Dugan:1990df,Herrlich:1994kh} affects the matching coefficients of physical operators only at the two-loop level.} We make use of the evanescent scheme of Refs.~\cite{Cirigliano:2020msr,Buhler:2023gsg}: the only evanescent bosonic operator is a total-derivative operator due to the Schouten identity
\begin{equation}
	\label{eq:SchoutenEvan}
	\E_S = \frac{1}{g_0^2} \left( \Box\tr[G_{\mu\nu}\widetilde G_{\mu\nu}] - 4 \p_\alpha \p_\mu \tr[ G_{\alpha\nu} \widetilde G_{\mu\nu} ] \right) \, ,
\end{equation}
whereas the fermionic evanescent operators are built from quark bilinears containing evanescent Dirac matrices,
\begin{equation}
	\label{eq:FermionicEvan}
	\E_i^{(n,m)} = \bar q \gamma_5 \hat\gamma_{\mu_1} \ldots \hat\gamma_{\mu_n} \gamma_{\nu_1} \ldots \gamma_{\nu_m} O^{F,i}_{\mu_1\ldots\mu_n\nu_1\ldots\nu_m}q \, ,
\end{equation}
where $O^{F,i}_{\mu_1\ldots\mu_n\nu_1\ldots\nu_m}$ consists of gauge fields and derivatives in $D$ dimensions. Due to the presence of explicitly evanescent indices on fields or derivatives, the projection onto the non-evanescent sector is efficiently done by keeping momenta and polarization vectors of external particles in $D=4$ space-time dimensions.

\section{An alternative scheme}
\label{eq:4DScheme}

The choice of MS renormalization scheme in the EFT affects the matching results. Although these scheme dependences need to drop out in relations between observables, certain scheme choices help to simplify the calculation. The scheme definitions of Ref.~\cite{Cirigliano:2020msr} are well suited for the present calculations in the $CP$-odd and flavor-conserving sector up to dimension six with only single operator insertions: the calculation in the HV scheme can be performed in $D$ dimensions, with the only explicit projection to 4 dimensions arising from the Levi-Civita symbol. In a more general context, these simplifications do not occur. Therefore, we also consider an alternative scheme choice, in which all non-evanescent higher-dimension operators are strictly kept in four space-time dimensions~\cite{Naterop:2023dek}, i.e., summed Lorentz indices in all operators only run over $\mu = 1, \ldots, 4$. We write the SFTE as
\begin{align}
	\label{eq:SFTE4DScheme}
	\O^R_{\widetilde G}(x,t) &= \sum_i \bar C_i(t,\mu) \bar\O_i^\mathrm{MS}(x,\mu) + \sum_i \bar C_{\N_i}(t,\mu)\bar\N_i^\mathrm{MS}(x,\mu) \nn
		&\quad + \sum_i \bar C_{\E_i}(t,\mu)\E_i^\mathrm{MS}(x,\mu) \,.
\end{align}
The last sum extends over evanescent bosonic operators that are not generated in the scheme of Sect.~\eqref{sec:Operators}: these bosonic evanescent operators can be written in terms of evanescent derivatives or gauge fields and the projection to the non-evanescent sector is again most easily done by keeping external momenta and polarization vectors in four space-time dimensions. Therefore, the projection onto the non-evanescent sector in terms of bare operators can be done in the same way in both schemes. The implications of the operator renormalization will be discussed in Sect.~\ref{sec:SchemeDependence}.


\begin{figure}[t]
	\centering
	\includegraphics[width=2.15cm]{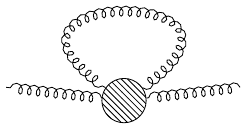} \;
	\includegraphics[width=2.15cm]{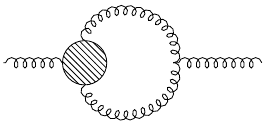} \;
	\includegraphics[width=2.15cm]{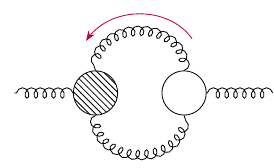} \;
	\includegraphics[height=2cm]{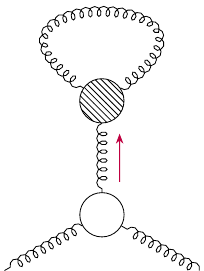}

	\caption{Feynman diagrams for the one-loop matching of the \cptgo{} to the QCD theta term. A filled blob denotes the insertion of the operator and an empty blob denotes a flow-time vertex. Crossed diagrams are not shown explicitly. The first and last diagrams vanish identically.}
	\label{fig:GGDiagrams}
\end{figure}

\begin{figure*}[t]
	\centering
	\mbox{}\\[-0.5cm]
	\includegraphics[height=1.65cm]{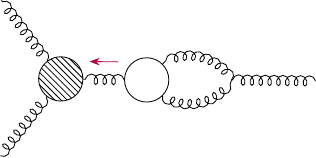} \,
	\includegraphics[height=1.65cm]{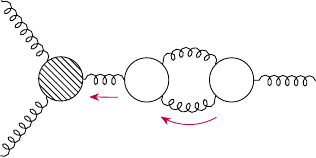} \,
	\includegraphics[height=1.65cm]{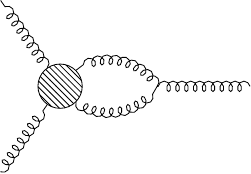} \,
	\includegraphics[height=1.65cm]{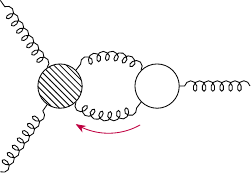} \,
	\includegraphics[height=1.65cm]{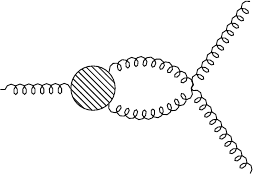} \,
	\includegraphics[height=1.65cm]{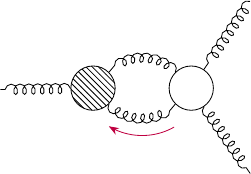} \\[0.2cm]
	\includegraphics[height=1.65cm]{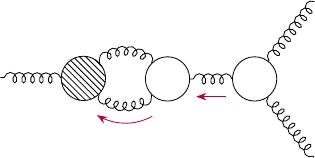} \;
	\includegraphics[height=1.65cm]{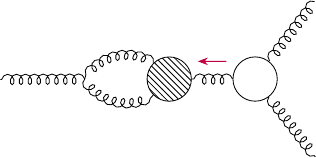} \;
	\includegraphics[height=1.65cm]{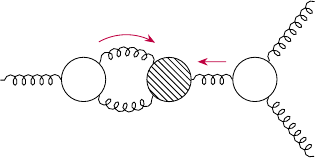} \;
	\includegraphics[height=1.65cm]{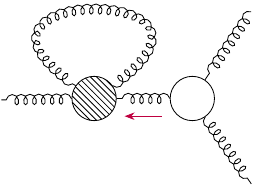} \;
	\includegraphics[height=1.65cm]{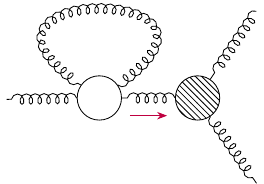} \\[0.2cm]
	\includegraphics[height=2.15cm]{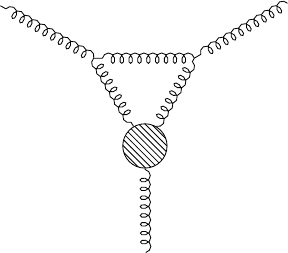} \,
	\includegraphics[height=2.15cm]{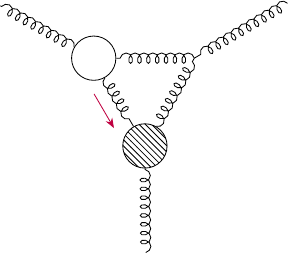} \,
	\includegraphics[height=2.15cm]{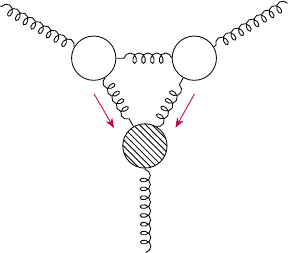} \,
	\includegraphics[height=2.15cm]{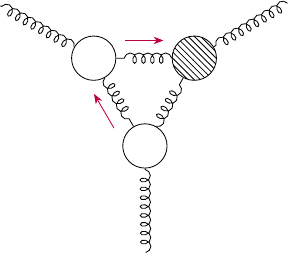} \,
	\includegraphics[height=2.15cm]{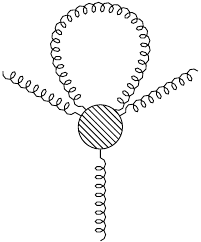} \,
	\includegraphics[height=2.15cm]{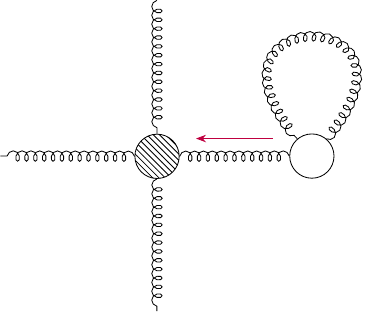} \,
	\includegraphics[height=2.15cm]{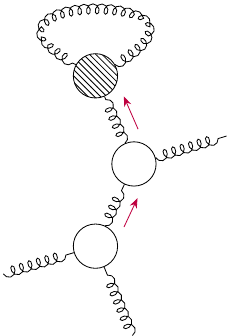}

	\caption{Feynman diagrams for the one-loop matching of the \cptgo{} to itself. Including crossed versions that are not listed explicitly, there are in total 56 diagrams.}
	\label{fig:GGGDiagrams}
\end{figure*}

\begin{figure}[t]
	\centering
	\includegraphics[height=2.15cm]{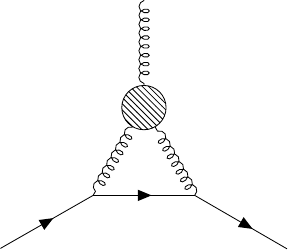} \qquad\qquad
	\includegraphics[height=2.15cm]{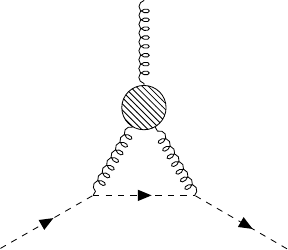}

	\caption{Left: Feynman diagram for the one-loop matching of the \cptgo{} to the qCEDM operator. Right: Feynman diagram contributing to the ghost-antighost-gluon three-point function.}
	\label{fig:qqGDiagrams}
\end{figure}

\begin{figure}[t]
	\centering
	\includegraphics[height=2.15cm]{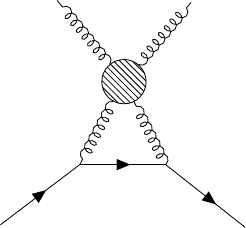} \;
	\includegraphics[height=2.15cm]{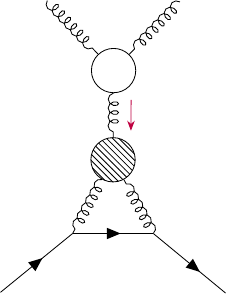} \;
	\includegraphics[height=2.15cm]{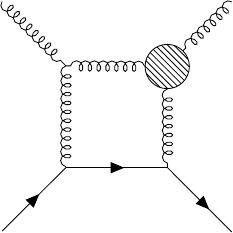} \\[0.2cm]
	\includegraphics[height=2.15cm]{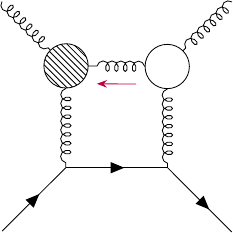} \;
	\includegraphics[height=2.15cm]{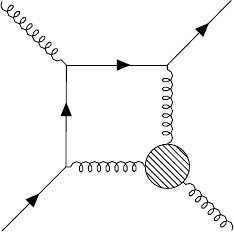} \;
	\includegraphics[height=2.15cm]{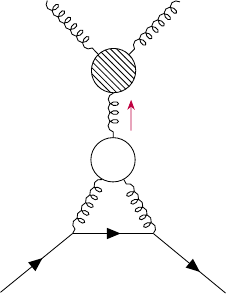}

	\caption{Feynman diagrams for the one-loop matching of the \cptgo{} to the nuisance operators. Diagrams with crossed gluons or reversed quark flow are not shown explicitly.}
	\label{fig:qqGGDiagrams}
\end{figure}

\section{Calculation}
\label{sec:Calculation}

The matching coefficients for the physical gluonic operators in the SFTE can be extracted from the gluon two- and three-point functions. In Figs.~\ref{fig:GGDiagrams} and~\ref{fig:GGGDiagrams}, we show the list of Feynman diagrams, omitting the ones related by crossing symmetry.  The matching to the quark chromo-EDM operator only requires the computation of a single diagram shown on the left of Fig.~\ref{fig:qqGDiagrams}. The determination of the full set of coefficients, including the coefficients of nuisance operators, also requires the calculation of the quark-antiquark-gluon-gluon four-point function, with diagrams shown in Fig.~\ref{fig:qqGGDiagrams}. For the extraction of the matching to the QCD theta term as well as to the remaining total-derivative operators, we insert momentum into the operator. Although only the matching coefficients of physical operators are of primary interest, we extract the complete set of coefficients in an off-shell matching: this provides an important consistency check of both our calculation and the completeness of the operator basis constructed in Ref.~\cite{Cirigliano:2020msr}. The ghost diagram shown on the right of Fig.~\ref{fig:qqGDiagrams} serves as a further cross check.

We compared multiple independent implementations of the entire loop calculation and we computed the matching for generic gauge parameters $\xi$ and $\alpha_0$, observing cancellation of the gauge-parameter dependence in the final results. Keeping generic gauge parameters increases the computational cost significantly and requires an efficient organization of the calculation, e.g., by making use of crossing symmetry in order to avoid that intermediate results become prohibitively large. To this end, we implemented our own routines in \texttt{Mathematica}.

Further, we checked that the $1/\varepsilon$ poles are cancelled by the counterterms that can be extracted from the known renormalization-group equations~\cite{Braaten:1990gq,Braaten:1990zt,Chang:1990dja,Jenkins:2017dyc,deVries:2019nsu}.

The matching calculation is greatly simplified by using the method of regions~\cite{Beneke:1997zp} with the inverse of the short flow time $\Lambda^2 = 1/t$ as the only hard scale.


\section{Results}
\label{sec:Results}

The matching of the \cptgo{} onto the theta term is power divergent,
\begin{equation}
	C_{\theta} = - \frac{9 \alpha_s C_A}{16\pi} \frac{1}{t} \, ,
\end{equation}
hence it necessitates a non-perturbative subtraction~\cite{Maiani:1991az,Kim:2021qae,Mereghetti:2021nkt}. This result supersedes the original finding of Ref.~\cite{Rizik:2020naq}. For the coefficients of the dimension-six operators, we obtain the following matching:
\begin{align}
	C_{\widetilde G} &= 1 + \frac{\alpha_s C_A}{12\pi}+\frac{3 \alpha_s C_A \log(8\pi\mu^2t)}{2 \pi} \, , \nn
	C_{CE} &= \frac{31 i \, \alpha_s C_A}{192\pi} + \frac{3 i \, \alpha_sC_A \log(8\pi\mu^2t)}{32\pi} \, , \nn
	C_{\p G} &= -\frac{179 \alpha_s C_A}{96 \pi } \, , \nn
	C_{\Box\theta} &= 0 \, .
\end{align}
The matching coefficients of the nuisance operators are given by
\begin{align}
	C_{\N_1} &= - \frac{5 i \alpha_s C_A}{192\pi}  \, , \nn
	C_{\N_8} &= - \frac{11 \alpha_s C_A}{384 \pi} - \frac{3 \alpha_s C_A \log(8\pi\mu^2t)}{64\pi}  \, , \nn
	C_{\N_{11}} &= - \frac{7 \alpha_s C_A}{128 \pi} - \frac{3 \alpha_s C_A \log(8\pi\mu^2t)}{64\pi}
\end{align}
and all other coefficients vanish at one loop. The Schouten-evanescent operator does not get a contribution either, $C_{\E_S}=0$. We do not list the coefficients of fermionic evanescent operators, as we simply project to the non-evanescent sector by keeping momenta and polarization vectors in four dimensions.

In the scheme where non-evanescent higher-dimension operators are kept in four space-time dimensions, we find the same power divergence, but slightly different results for the finite pieces of the matching coefficients of the dimension-six operators:
\begin{align}
	\bar C_{\widetilde G} &= 1 + \frac{3 \alpha_s C_A \log(8\pi\mu^2t)}{2 \pi} \, , \nn
	\bar C_{CE} &= \frac{11 i \, \alpha_s C_A}{64\pi} + \frac{3 i \, \alpha_sC_A \log(8\pi\mu^2t)}{32\pi} \, , \nn
	\bar C_{\p G} &= -\frac{61 \alpha_s C_A}{32 \pi } \, , \nn
	\bar C_{\Box\theta} &= 0 \, , \nn
	\bar C_{\N_1} &= - \frac{i \alpha_s C_A}{64\pi}  \, , \nn
	\bar C_{\N_8} &= - \frac{7 \alpha_s C_A}{128 \pi} - \frac{3 \alpha_s C_A \log(8\pi\mu^2t)}{64\pi}  \, , \nn
	\bar C_{\N_{11}} &= - \frac{9 \alpha_s C_A}{128 \pi} - \frac{3 \alpha_s C_A \log(8\pi\mu^2t)}{64\pi}
\end{align}
and vanishing one-loop coefficients of the remaining nuisance operators.

\section{Scheme dependence and evanescent operators}
\label{sec:SchemeDependence}

The calculation in the scheme of Sect.~\ref{eq:4DScheme} is most easily done by replacing the flowed operator on the LHS of Eq.~\eqref{eq:SFTE4DScheme} by the flowed operator in four space-time dimensions,
\begin{equation}
	\label{eq:FlowedCP3GO4D}
	\bar \O^R_{\widetilde G}(x,t) \coloneqq \frac{1}{g^2} \overline{\tr[G_{\mu\nu} G_{\nu\lambda} \widetilde G_{\lambda\mu}]} \, .
\end{equation}
In this way, one can also avoid a tree-level matching contribution to evanescent operators in this scheme. However, since the flowed operator is UV finite, this choice cannot affect the matching coefficients of non-evanescent operators, $\bar C_i$ and $\bar C_{\N_i}$, which we demonstrate explicitly in the following.

If we perform the matching using the original flowed operator~\eqref{eq:FlowedCP3GO}, one might expect to find exactly the same matching coefficients of non-evanescent operators as in the first scheme, since the operators in the second scheme only differ by evanescent contributions. However, this argument neglects the renormalization of the operators. Indeed, when using the flowed \cptgo{}~\eqref{eq:FlowedCP3GO} in the matching~\eqref{eq:SFTE4DScheme}, we already get a tree-level contribution to the evanescent operator
\begin{equation}
	\label{eq:BareEvanescent}
	\E_{\widetilde G} = \frac{1}{g_0^2} \tr[G_{\mu\hat\nu} G_{\hat\nu\lambda} \widetilde G_{\lambda\mu}] = \O_{\widetilde G} - \bar\O_{\widetilde G} \, ,
\end{equation}
with matching coefficient $\bar C_{\E_{\widetilde G}}^\text{tree} = 1$. The usual way to renormalize evanescent operators is not by pure subtraction of $1/\varepsilon$ poles, but one should rather include finite physical counter\-terms that compensate the finite non-evanescent contributions arising from the insertion of the evanescent operators into loop diagrams~\cite{Dugan:1990df,Herrlich:1994kh}, imposing\footnote{Despite the finite renormalization, we denote the renormalized evanescent operators by $\E^\mathrm{MS}_i$, since this scheme is indeed minimal~\cite{Dugan:1990df}. See also Refs.~\cite{Dekens:2019ept,Aebischer:2022aze,Fuentes-Martin:2022vvu} for recent discussions of evanescent operators in matching calculations.}
\begin{equation}
	\label{eq:EvanescentRenormalizationCondition}
	\< \E_i^\mathrm{MS} \>\big|_\text{phys} = 0 \, .
\end{equation}
Relating renormalized to bare operators by
\begin{equation}
	\E_i^\mathrm{MS} = (\delta_{ij} + \Delta_{ij}^{\E\E}) \E_j + \Delta_{ij}^{\E\O} \O_j + \Delta_{ij}^{\E\N} \N_j \, ,
\end{equation}
this implies at one loop
\begin{equation}
	0 = \< \E_i \>\big|^{1L}_\text{phys} + \Delta_{ij}^{\E\O} \< \O_j \> \big|_\text{phys}^\text{tree} + \Delta_{ij}^{\E\N} \< \N_j \> \big|_\text{phys}^\text{tree} \, .
\end{equation}
An explicit calculation of the insertions of the evanescent operator~\eqref{eq:BareEvanescent} leads to the finite renormalizations
\begin{align}
	\label{eq:FiniteRenormalizations}
	\Delta^{\E\O}_{\widetilde G, \widetilde G} &= \frac{\alpha_s C_A}{12\pi} \, , \quad
	\Delta^{\E\O}_{\widetilde G, CE} = -\frac{i \alpha_s C_A}{96\pi} \, , \quad
	\Delta^{\E\O}_{\widetilde G, \p G} = \frac{\alpha_s C_A}{24\pi} \, , \nn
	\Delta^{\E\N}_{\widetilde G, 1} &= -\frac{i \alpha_s C_A}{96\pi} \, , \quad
	\Delta^{\E\N}_{\widetilde G, 8} = \frac{5\alpha_s C_A}{192\pi} \, , \quad
	\Delta^{\E\N}_{\widetilde G, 11} = \frac{\alpha_s C_A}{64\pi} \, .
\end{align}
Writing the matching coefficients as $\bar C_i = \bar C_i^\text{tree} + \bar C_i^{1L}$, expanding the SFTE~\eqref{eq:SFTE4DScheme} in terms of bare operators, and applying the method of regions gives at one loop
\begin{equation}
	\< \O^R_{\widetilde G}(t) \> \big|^{1L}_\text{hard,finite} = \left( \bar C_i^{1L} + \Delta^{\E\O}_{\widetilde G,i} \right) \< \O_i\>\big|^\text{tree} + \bar C_{\E_i}^{1L} \< \E_i \>\big|^\text{tree} \,,
\end{equation}
where for simplicity we suppressed field-renormalization factors and included nuisance operators in the sum over physical operators $\O_i$. Instead, if we write the SFTE for the flowed \cptgo{} in four dimensions~\eqref{eq:FlowedCP3GO4D}, there is no tree-level matching contribution to evanescents, in particular $\bar C_{\E_{\widetilde G}}^\text{tree}{}' = 0$, and the matching equation reads
\begin{equation}
	\< \bar\O^R_{\widetilde G}(t) \> \big|^{1L}_\text{hard,finite} = \bar C_i^{1L} \< \O_i\>\big|^\text{tree} + \bar C_{\E_i}^{1L}{}' \< \E_i \>\big|^\text{tree} \,.
\end{equation}
Therefore, while the matching coefficients of the evanescent operators do change, the choice of taking either $\O^R_{\widetilde G}(x,t)$ or $\bar\O^R_{\widetilde G}(x,t)$ as the flowed operator indeed leads to the same matching coefficients of physical (and nuisance) operators, since the finite renormalizations~\eqref{eq:FiniteRenormalizations} exactly reproduce the difference in the coefficient of physical structures between the insertions of $\O^R_{\widetilde G}(x,t)$ and $\bar\O^R_{\widetilde G}(x,t)$. The calculation therefore confirms the expectation that the matching coefficients depend on the dimensional scheme for the MS operators (as illustrated by the difference between the matching coefficients $C_i$ and $\bar C_i$), but the dimensional scheme for the flowed operators does not affect the matching coefficients of the non-evanescent sector, since the operators are already regulated by the flow. Importantly, this only works if the evanescent operators are correctly renormalized, i.e., by imposing the condition~\eqref{eq:EvanescentRenormalizationCondition}, which separates the physical from the evanescent sectors.


\section{Conclusions}
\label{sec:Conclusions}

We have presented the results for the one-loop matching of the $CP$-violating three-gluon operator between the gradient flow and MS schemes. These results complete the one-loop matching of the $CP$-violating and flavor-conserving sector of the LEFT to the gradient flow, which is the sector of the theory relevant for the neutron EDM. We have provided the results for two different renormalization schemes on the MS side, which are both based on the 't~Hooft--Veltman scheme but use a different operator basis. We have checked that the matching results are independent of the dimensional scheme used to define the flowed operator. This requires a correct renormalization of evanescent operators, following the scheme introduced in Refs.~\cite{Dugan:1990df,Herrlich:1994kh}. Interestingly, in the scheme where physical effective operators are kept in four space-time dimensions, the one-loop contribution to the self-matching of the \cptgo{} is purely logarithmic and vanishes at the matching scale $\mu = (8\pi t)^{-1/2}$. In the scheme of Ref.~\cite{Cirigliano:2020msr}, we do find a finite one-loop self-matching contribution. The perturbative uncertainty of the one-loop gradient-flow matching has been studied previously~\cite{Mereghetti:2021nkt,Buhler:2023gsg}: in the context of the neutron EDM, the target of a $10-25\%$ uncertainty on hadronic matrix elements~\cite{Alarcon:2022ero} as well as the experimental prospects for improved sensitivities~\cite{Ito:2017ywc,nEDM:2019qgk,Wurm:2019yfj,Martin:2020lbx,n2EDM:2021yah} provide motivation to extend these calculations in the future to the two-loop level~\cite{Artz:2019bpr,Harlander:2022tgk,Harlander:2022vgf}.

\section*{Acknowledgments}
We are grateful to E.~Mereghetti for collaboration at an early stage of the project and for valuable comments on the manuscript. We further thank J.~B\"uhler and L.~Naterop for useful discussions.
OLC and PS gratefully acknowledge financial support by the Swiss National Science Foundation (Project No.~PCEFP2\_194272).
CJM is supported in part by USDOE grant No.~DE-SC0023047.
AS acknowledges funding support under the 
National Science Foundation grant PHY-2209185.
AS thanks R. Harlander for discussions 
on the gradient flow in perturbation theory.


\appendix

\nocite{*}

\bibliographystyle{apsrev4-1_mod}
\balance
\biboptions{sort&compress}
\bibliography{Literature}

\end{document}